\documentclass[10pt]{iopart}
\begin{document}
\title[Frenet-Serret dynamics]{Frenet-Serret
dynamics}

\author{G Arreaga\dag ,
R Capovilla\dag and
J Guven\ddag}
\address{\dag\
Departamento de F\'{\i}sica,
Centro de Investigaci\'on y de Estudios Avanzados del IPN, 
 Apdo. Postal 14-740,07000 M\'exico, DF, MEXICO}
\address{ \ddag\ Instituto de Ciencias Nucleares,
Universidad Nacional Aut\'onoma de M\'exico,
 Apdo. Postal 70-543, 04510 M\'exico, DF, MEXICO}

\begin{abstract}
We consider the motion of a particle 
described by an action that is a functional 
of the Frenet-Serret [FS] curvatures
associated with the embedding of its worldline 
in Minkowski space. We develop a  theory of deformations 
tailored to the FS frame. 
Both the Euler-Lagrange equations and
the physical invariants of the motion associated with the 
Poincar\'e symmetry of Minkowski space, the mass and the spin of the particle, 
are expressed in a simple way in terms of these curvatures. 
The simplest non-trivial model of this 
form, with the Lagrangian depending on 
the first FS (or geodesic) curvature, is integrable. 
We show how this integrability can be deduced from the 
Poincar\'e invariants of the motion. 
We go on to explore the structure of these invariants 
in higher-order models.
In particular, the integrability of the
model described by a Lagrangian that is a 
function of the second FS 
curvature (or torsion) is established in a 
three dimensional ambient spacetime.
\end{abstract}


\pacs{1480P,0240H,0425,0450}

\section{Introduction}

The simplest action describing the motion of a particle 
is proportional to the proper time along  the trajectory
of the particle in spacetime, or worldline. When 
there are no external fields, this is the unique Poincar\'e invariant 
action that can be constructed using only the velocity of the particle.

In this paper,  we examine a hierarchy of geometrical models
describing particle motion that involve
successively higher derivatives of the embedding functions
of the worldline \cite{pisa,zoller,Ply,Nes,KP1,KP2,IIT,RR, Nerse}. Such
models
arise as a description of an
isolated object when its internal structure  is not resolved. 
Indeed, the effective bosonic theory describing a supersymmetric
particle when fermionic degrees of freedom have been integrated out
can itself be cast as a non-local theory of this form \cite{Poly}.
We will focus, however, not so much on any specific model, but 
on  the relationship between the order of 
a given local geometrical action and the dimension 
of the background geometry.  While this point is trivial
for a free particle,
it is not so for higher order models.

Let $N+1$ be the dimension of our spacetime.
Suppose also that the trajectory is sufficiently smooth so that 
the Frenet-Serret [FS] frame which is adapted to it 
\cite{DHOT,Nesto} is defined, namely, the velocity $V$ along the
trajectory is 
differentiable $N$ times. We can then associate a 
 curvature $\kappa_{n}$ with each of the first $N$ derivatives of $V$:
$\kappa_1$ is the acceleration of the particle; $\kappa_2$ 
describes the torsion of its trajectory 
and so on. Once the FS curvatures have been specified 
the course of the particle can  be reconstructed up to a Poincar\'e
rigid motion 
(as in \cite{Spivak}, Corollary 7.4).
In this way the frame provides a complete 
kinematical description of the motion.
Every geometrical scalar associated with the trajectory can be  
constructed using only the FS curvatures and their derivatives.
We thus abandon the embedding functions as our 
primary variables; their derivatives replaced by geometrically meaningful 
objects and
cast the action as a functional of 
the FS curvatures, 
\[
S=\int d\tau
L ( \kappa_1,\kappa_2,\cdots,  \kappa_1', \cdots ) \;,
\]
where $\tau$ is the proper time along the trajectory of the
particle and the prime denotes  a derivative with respect to $\tau$.

In its original Euclidean incarnation, 
the Lagrangian $L(\kappa_1)$ can be traced 
back to the beginning of the calculus of variations
in Euler's description of space curves (see {\it e.g.} \cite{Giaquinta}). 
The model is exactly solvable, a fact not at all obvious if the action is 
expressed in terms of its original embedding variables. 
By adapting the differential geometry of Pfaffian forms
to the FS frame, however,
Dereli {\it et al.} were able to show that the relativistic particle model
was integrable by expressing $\kappa_1$
as  a quadrature \cite{DHOT}. An 
alternative approach without Pfaffian forms
was developed recently by Nesterenko {\it et al.} 
\cite{Nesto}. 

Beyond the first curvature, there has been very little progress made to date. This is 
understandable:  the equations of motion are an intractable muddle
when expressed directly in terms of the embedding functions; 
the Hamiltonian formalism based on them, preferred by physicists with
their 
sight set on quantization, even worse. To some extent, we are able to
overcome this obstacle by developing 
a theory of worldline deformations which is tailored  to the 
FS frame: the deformation in the worldline embedding functions
is decomposed with respect to this frame;
the infinitesimal change induced in the 
curvatures under this deformation is cast as a linear differential operator acting on these 
projections. 
The expressions themselves turn out to be far simpler than the 
corresponding expressions for the deformation of the relevant order of
derivatives of the embedding functions. The FS frame projections 
also provide a natural set of independent 
deformations in the variational principle; the 
projections of the Euler-Lagrange equations along these directions 
assume a strikingly simple form.

The early work on $L(\kappa_1)$ demonstrated very clearly the natural division of  
the dynamical problem into two parts: 
in the first part two first integrals of the equations of motion 
in terms of the curvature of the worldline and its first derivative 
are identified and combined to provide a quadrature; in 
the second part the FS equations are integrated 
for the position vector of the worldline 
of the particle for the given $\kappa_1$. 
In higher order systems, 
it is not easy to identify first integrals.
We are guided, however, by 
Nesterenko {\it et al.} who 
recognized in the context of $L(\kappa_1)$, that these first integrals 
are none other than the Casimir 
invariants of the motion which are a consequence of the Poincar\'e symmetry 
of Minkowski space, the mass $M^2$ and  
the spin $S^2$, of the particle associated with the 
conserved linear and angular momenta respectively.

We will employ Noether's theorem to determine the Poincar\'e invariants 
for any local Lagrangian constructed from the FS curvatures.
Thus, there always exist two non-trivial first integrals of the equations of motion.
Using the deformation theory we have developed, these invariants  
can be cast  in a simple as well as instructive
way in terms of the FS curvatures. 

In the case of $L(\kappa_1)$, the conservation of angular momentum identifies
$\kappa_2$ as a function of $\kappa_1$. This permits us
to cast the second integral of the motion as a quadrature,
$
\kappa_1{}'{}^2  + V( \kappa_1 ; S^2 , M^2 ) =0\,,
$
which mimics the radial motion of a fictitious
non-relativistic particle  in a central potential on the normal plane. 

In general, the system of equations will not be completely integrable:
the first integrals do not provide a quadrature. This appears to be the 
case for $L(\kappa_1,\kappa_2)$.
Remarkably, however, we show that $L(\kappa_2)$, like the 
lower order $L(\kappa_1)$, is completely integrable when $N=2$
(the analogous result in three-dimensional Euclidean space
will be considered elsewhere). This time,
however, the integrals of motion are
entangled in a rather less trivial way. Nonetheless, 
the first integral of the equations of motion can be cast as
a quadrature for $\kappa_2$,
$
\kappa_2{}'{}^2 + V(\kappa_2 ; S^2 , M^2 ) =0\,,
$
and $\kappa_1$ as function of
$ (\kappa_2 , S^2 , M^2 )$, thus once again permitting one to determine the curve 
up to a rigid motion. 
The equations of motion never need to be integrated explicitly.

The paper is organized as follows:
In Sect. 2. we summarize the relevant 
worldline geometry. 
In Sect. 3. we develop the theory of deformations of an arbitrary worldline.
In Sect. 4., we examine general features of relativistic 
particle described by any (local) Lagrangian of the form
$L ( \kappa_1,\kappa_2,\cdots,  \kappa_1', \cdots )$. 
In Sect. 5  we specialize to  a Lagrangian,
$L=  L (\kappa_1)$. 
In Sect. 6
we extend our treatment to models which depend 
on the second curvature, $L = L ( \kappa_2 )$ .

\section{Worldline geometry}

In this section we consider the geometry of embedded 
timelike curves from two different points of view. First we use a 
language which is the
one-dimensional version of the geometry of higher dimensional
embedded surfaces in an arbitrary parametrization.
Next we consider the description in terms of the FS basis for the curve,
which exploits the existence of 
a preferred parameter, proper time.
Finally we show how these two descriptions are related.
To our knowledge, this relationship has not 
been explored before. We demonstrate that, 
whereas the higher order FS curvatures 
assume an impenetrable form when expressed in terms of derivatives
of the 
embedding functions, they assume a remarkably simple 
form in terms of the unit $k^i$ and their 
derivatives. Indeed, 
the top FS curvature is a simple polynomial in these variables.

We consider a relativistic particle whose trajectory is described by the 
worldline, $x^{\mu}=X^{\mu}(\xi) $,
 embedded in Minkowski spacetime 
$\{ M,\eta_{\mu\nu}\}$ of dimension $N+1$ (we use a signature
with only one minus sign; $\mu,\nu, \cdots = 0,1,\cdots, N$). $\xi$ is an 
arbitrary parameter. The vector tangent to the curve is $\dot X^\mu$, 
where the overdot represents a derivative with respect to the parameter $\xi$.
The one-dimensional induced metric on the curve is
$\gamma = \dot{X}^\mu \dot{X}^\nu \eta_{\mu\nu}
=\dot{X} \cdot \dot{X}$.
The infinitesimal proper time elapsed along the trajectory is then given by
$d\tau = \sqrt{- \gamma} d\xi$. 
If one uses  proper time as the parameter
(taking advantage of the
fact that the intrinsic geometry of the 
curve is trivial)  then
$\gamma = - 1$. 

In our first description of the geometry,
the normal frame is not fixed: an arbitrary set 
of vectors normal to the curve, $n^\mu{}_i$, can be
defined implicitly by $n_i \cdot  \dot{X} = 0$,
and normalized $ n_i \cdot n_j  = \delta_{ij} $
($i,j \dots = 1,2,\cdots, N$). Normal indices are 
raised and lowered with the Kronecker delta.
The one-dimensional analogue of the Gauss-Weingarten 
equations is given by 
\begin{equation}
\ddot{X} = \Gamma \dot{X} - K^i n_i\,, 
\;\;\;\;\;\;\;\;\;
\dot{n}_i = k_i \dot{X} + \omega_i{}^j n_j\,.
\label{eq:gw}
\end{equation}
Here we have introduced the one-dimensional affine connection
$\Gamma =  \gamma^{-1} \dot{X} \cdot  \ddot{X} = 
(1/2) \gamma^{-1} \dot\gamma $.
We define the covariant derivative under worldline reparametrizations
$ \nabla = (d/d\xi ) - \Gamma$. The connection $\Gamma$ 
vanishes in the parametrization by proper time.
The extrinsic curvature along the $i$-th normal is 
$K_i = - n_{i} \cdot \ddot{X}$,
and the point-like analogue of the mean extrinsic
curvature is the scalar under reparametrizations $ k_i = \gamma^{-1} K_i $.
The $K_i$ are genuine curvatures
involving second derivatives of the embedding,
as distinct from the Frenet-Serret curvatures 
which we discuss below.

The arbitrariness in the choice of frame does however come at a price -
an additional
one-dimensional
connection needs to be introduced to maintain covariance 
under normal rotations. 
The extrinsic twist, or normal form,  
$\omega_{ij} = n_{j} \cdot \dot{n}_i$ is the connection
associated with the freedom of rotations of the normals. We use it to define 
the covariant derivative under rotation of the normals
$\tilde\nabla = \nabla - \omega$. We will also use the symbol
$\tilde{D}$  to denote the covariant derivative $\tilde\nabla$
using proper time as the parameter.

The existence of 
a natural parameter, proper time, leads to an alternative
description, unique to curves, which is the generalization
to higher ambient space dimensions of the classical 
FS formalism for space curves (see {\it e.g.}
\cite{doCarmo}, \cite{Spivak} ).
In this approach, one introduces an orthonormal basis 
$\{ X' , \eta_i \} $, where a prime denotes derivative
with respect to proper time. The FS equations
for a curve in $N+1$ dimensions are 
\begin{eqnarray}
X'' &=& \kappa_1 \eta_1 \,, 
\nonumber
\\
\eta_1{}' &=& \kappa_1 X' - \kappa_2 \eta_2 \,, 
\nonumber \\
\eta_2{}' &=& \kappa_2 \eta_1 - \kappa_3 \eta_3 \,, 
\nonumber \\
\dots      &{}& \dots  \label{eq:frenet} \\
\eta_{N-1}{}' &=& \kappa_{N-1} \eta_{N-2} - \kappa_N \eta_N \,, 
\nonumber\\
\eta_N{}' &=& \kappa_N \eta_{N-1} \,.
\nonumber
\end{eqnarray}
where $\kappa_i$ represents the $i^{\rm th}$ FS
curvature. In an ambient space of dimension $N+1$, there
are at most $N$ FS curvatures. $\kappa_1$ is the
geodesic curvature. Implicit in this definition is that the embedding functions
$X^\mu$ are $N+1$ times differentiable, and that the $\kappa_i$
do not vanish. If they vanish only at isolated points however
the FS basis can still be defined consistently 
(see \cite{Spivak}, Ch. 7.)

There are some important results of the geometry of curves
that we will use extensively in this paper. First, the
fundamental theorem for curves says that
for a curve embedded in an $N+1$-dimensional 
space its $N$ FS curvatures determine the curve,
up to rigid motions. The actual trajectory can always
be obtained from the curvatures by quadratures. This implies
that
the curvatures can be used as  a set of natural auxiliary variables
for the description of curves. In addition,
if the $i-$th FS
curvature vanishes, so do all the higher ones and,  moreover, 
the curve will lie in an $i$-dimensional subspace. 
 For example, if $\kappa_2 = 0$ then the motion lies on a Minkowski plane,
if $\kappa_3 = 0$ the motion is in $2+1$ dimensions.
(For the Euclidean case see \cite{Spivak}, Th. 7.5.)
Let us also remark that the top FS curvature 
$\kappa_N$
has a special status. It is possible to express its content in a way
which depends on the orientation of the ambient space. This
{\it signed} curvature will be denoted with an overbar,
$\overline{\kappa}_N $, so that (see below for an explicit 
definition of $\overline\kappa_N$, valid
for any $N$)
\begin{equation}
\overline{\kappa}_N^2 =
\kappa_N^2\,.
\label{eq:signed}
\end{equation}

Let us turn our attention now to the problem of 
relating these two descriptions of the geometry
of curves. We will express the FS
curvatures in terms of $k^i$ and its derivatives.
In addition, we obtain an explicit expression for
the curvature dependent rotation matrix taking 
an  arbitrary normal basis $n_i$ into 
FS normal form $\eta_i$: 
\begin{equation}
\eta_i = R_i{}^j \; n_j\,.
\label{eq:rot}
\end{equation}
We begin by parametrizing Eq.(\ref{eq:gw}) by proper time:
\begin{equation}
X'' =  k^i \; n_i\,, 
\;\;\;\;\;\;\;\;\;\;\
\tilde{D} n_i = k_i \; X ' \,,
\label{eq:gwp}
\end{equation}
where the symbol $\tilde{D}$ was defined below Eq.(\ref{eq:gw}).
Now, comparison of the first with the first of  (\ref{eq:frenet}) gives
immediately $\kappa_1 \eta_1 = k^i n_i $. It follows that
the first FS curvature is simply the modulus of
$k^i$,
$\kappa_1 = k = \sqrt{k^i k_i }$.
We have also that
$\eta_1 = \hat{k}^i n_i $,
where we have introduced the unit vector $\hat{k}^i = k^i / k $, 
so that we find $ R_1{}^i = \hat{k}^i$.
For the next order, we take a derivative of 
$\eta_1$, and
using (\ref{eq:gwp}), we obtain
$
\eta_1{}' = ( \tilde{D} \hat{k}^i ) n_i
+ \kappa_1 X' $.
Comparison with the second FS equation  gives
\begin{equation}
\kappa_2 = \sqrt{ \tilde{D} \hat{k}^i 
\tilde{D} \hat{k}_i }\,,
\label{eq:kappa2}
\end{equation}
and $ \eta_2 = - \kappa_2{}^{-1} ( \tilde{D} \hat{k}^i )
n_i $,
so that $R_2{}^i =  -\kappa_2^{-1}  \tilde{D} \hat{k}^i  $.  
Note that $\eta_1 \cdot \eta_2 = 0$ follows from the
unit vector fact $ \hat{k}_i \tilde{D} \hat{k}^i = 0 $.
The interesting non-obvious fact is that the 
second FS curvature is simply the modulus of $\tilde D \hat{k}^i$.
We can continue in the same fashion for the higher
FS curvatures. One finds that $\kappa_n$ is proportional to the 
modulus of the projection of $\tilde D^{n-1} \hat k^i$ 
orthogonal to all lower derivatives of $\hat k^i$. 
What becomes clear is that the $N-1$ derivatives of
$k^i$, together with $k^i$ itself provide a basis for the
normal directions.  This implies that one can construct a set
of  normal forms that encode completely the information 
contained in the higher FS curvatures: 
\begin{equation}
\fl
T^{(N)}_{i_1\cdots i_{N-r}} =  {1 \over \sqrt{(N-r)!}} 
\; {1\over k^{r-1}}\varepsilon_{i_1\cdots i_{N-r} 
j_1\cdots j_{r}}
k^{j_1} (\tilde{D} k^{j_2}) \cdots (\tilde{D}^{r-1} k^{j_r}) \,,
\label{eq:tis}
\end{equation}
where, $r=1,\cdots N$. 
Here 
$ \varepsilon_{i_1 \cdots i_N } =
\varepsilon_{\alpha_1 \cdots \alpha_N \mu}
n^{\alpha_1}{}_{i_1} \cdots n^{\alpha_N}{}_{i_N} {X}^\mu{}'$ 
is the normal Levi-Civita density, with $\varepsilon_{\alpha_1 \cdots \alpha_N \mu}$ the Minkowski background Levi-Civita density.
Note that we have the alternative expression in terms of the unit
$\hat{k}^i$ and
its derivatives, 
\begin{equation}
T^{(N)}_{i_1\cdots i_{N-r}} = {1 \over \sqrt{(N-r )!}}\; k \; \varepsilon_{i_1\cdots i_{N-r} j_1\cdots j_{r}}\;
\hat{k}^{j_1} \tilde{D} \hat{k}^{j_2}\cdots (\tilde{D}^{r-1} \hat 
k^{j_r})\,.
\end{equation}
For example, for a space curve we have 
$T^{(2)}_i = k \; \varepsilon_{ij}\; \hat{k}^j $, and
$T^{(2)} = k \; \varepsilon_{ij}\; \hat{k}^i \; \tilde{D} \hat{k}^{j}$,
whereas for a curve in four dimensions,
\begin{equation}
\fl
T^{(3)}_{ij} = {k \over \sqrt{2}} \; \varepsilon_{ijl} \; \hat{k}^l \,, \;\;\;\;\;
T^{(3)}_i = k \;\varepsilon_{ijl} \;\hat{k}^j \; \tilde{D} \hat{k}^{l}\,,
\;\;\;\;\;
T^{(3)} = k \;\varepsilon_{ijl}\; \hat{k}^i \; (\tilde{D} \hat{k}^{j} ) \; (\tilde{D}^2 \hat{k}^{l} )\,.
\end{equation}
These normal forms contain the same information as the 
FS curvatures, in the sense that the following relation
holds,
\begin{equation}
T^{(N)}_{i_1\cdots i_{N-r}} T^{(N)}{}^{i_1\cdots i_{N-r}} 
= \kappa_1^2 \kappa_2^2 \cdots \kappa_r^2\,.
\label{eq:tcu}
\end{equation}
For the signed top FS curvature we find
the relation
 \begin{equation}
 T^{(N)}  = \kappa_1 \kappa_2 \cdots
\kappa_{N-1} \overline{\kappa}_N \,,
\end{equation}
which provides a simple definition for $\overline{\kappa}_N $ valid for any $N$ 
that satisfies  (\ref{eq:signed}). The expression
for $\overline{\kappa}_N$ directly in terms of
the embedding functions is much more complicated. We note, in particular, that
the (signed) torsion of the worldline 
(a spacetime pseudoscalar) is defined with respect to the embedding functions by
\begin{equation}
\overline{\kappa}_2 = { \varepsilon_{\mu \nu \rho} \dot{X}^{\mu} \ddot{X}^{\nu} 
 \mathop{X}\limits^{\ldots}{}^\rho \over 
 \dot{X}^2 \ddot{X}^2 - (\dot{X} \cdot \ddot{X} )^2 }\,,
\end{equation}
whereas in terms of $\hat k^i$ and its derivative it assumes the remarkably simple polynomial form, $
\overline{\kappa}_2 =  \varepsilon_{i j} \; \hat k^i \;
\tilde{D} \hat k^j$.

An alternative approach is the following: let us introduce the
objects $\Pi_{(n+1)}{}^{ij}$, defined recursively by ($n \le N$)
\begin{equation}
\fl
\Pi_{(n+1)}{}^{ij} = 
\Pi_{(n)}{}^{ij} - {1 \over \kappa_{n}^2\cdots \kappa_1^2 }
\; \Pi_{(n)}{}^{i}{}_k 
\; \Pi_{(n)}{}^{j}{}_l  
\;  ( \tilde{D}^{n-1} k^k ) \; (\tilde{D}^{n-1} k^l ) \,,
\end{equation}
with $\Pi_{(1)}{}^{ij} = \delta^{ij}$.  
The lowest cases are
\[
\Pi_{(2)}{}^i{}_j  = \delta^{i}{}_j - \hat{k}^i \; \hat{k}_j \,,
\;\;\;\;\;\;\;
\Pi_{(3)}{}^i{}_j  = \delta^{i}{}_j - \hat{k}^i \; \hat{k}_j 
- {1 \over \kappa_2^2} (\tilde{D} \hat{k}^i )\; (\tilde{D} \hat{k}_j )\,. 
\]
These objects are projection matrices,
\begin{equation}
\Pi_{(n)}{}^i{}_j \; \Pi_{(n)}{}^j{}_k
= \Pi_{(n)}{}^i{}_k\,, \;\;\;\;\;\;\;\; 
 \Pi_{(n)}{}^i{}_i = N-n + 1\,,
\label{eq:pro}
\end{equation}
which satisfy 
\begin{equation}
\Pi_{(n)}{}^i{}_j \; \Pi_{(m)}{}^j{}_k
= \Pi_{(n)}{}^i{}_k\,, \;\;\;\; 
\Pi_{(n)}{}^i{}_j \; \tilde{D}^{m-1} k^j = 0\,.
\;\;\;\; ( n > m ) 
\label{eq:nm}
\end{equation}
In general, we have the following relation with the normal forms 
introduced earlier,
\begin{equation}
\Pi_{(n)}{}^{i}{}_j = {(N-n+1) \over \kappa_1^2 \cdots
\kappa_{n-1}^2} \;\;
T^{(N)}{}^{i\, i_1\cdots
i_{N-n}} T^{(N)}{}_{j \, i_1  \cdots i_{N-n}}\,.
\label{eq:genx}
\end{equation} 
With the help of these projectors, we find that 
the FS curvatures are defined recursively by
\begin{equation} 
\kappa_{n} = {1 \over  \kappa_{n-1} \kappa_{n-2} \cdots \kappa_1 }
\sqrt{( \tilde{D}^{n-1} k_i )\;  \Pi_{(n)}{}^{ij} \; 
 (\tilde{D}^{n-1} k_j ) }\,. 
\label{eq:kappan}
\end{equation}
Using (\ref{eq:genx}), it is easy to show that this expression agrees
with what we have found earlier, (\ref{eq:tcu}).
For $\kappa_{2}$  it reduces to the
expression (\ref{eq:kappa2}). The next FS curvature can be
written as
\begin{equation}
\kappa_3 = {1 \over \kappa_2 } \sqrt{
(\tilde{D}^2 \hat{k}^i)(\tilde{D}^2 \hat{k}_i) - (\kappa_2{}')^{2} - \kappa_2^4 }\,.
\label{eq:kappa3}
\end{equation}
We emphasize that this is a considerable improvement on the  corresponding expression
directly in terms of derivatives of the embedding functions.

We also obtain the sought for general expression 
(up to a sign) for the rotation matrices
defined by (\ref{eq:rot}), as
\begin{equation} 
R_i{}^j = {1 \over \kappa_{i} \kappa_{i-1} \cdots
\kappa_{1}} \;
\Pi_{(i)}{}^{jl} \; \tilde{D}^{i-1} k_l\,.
\end{equation}

To conclude this section we note that the expressions we have derived 
generalize in a straightforward way  to the case of an arbitrary curved background spacetime
$\{ M , g_{\mu\nu} \}$. We would have
$\eta_{\mu\nu} \to g_{\mu\nu}$ and derivation by proper time becomes
the covariant derivative along the tangent vector to the wordline.
In particular, this changes  the definition of $k^i$, which 
depends now on the background covariant derivative. 

\section{Worldline deformations}

In this section we analyze the change in the geometry of
the worldline due to an infinitesimal deformation 
$X^\mu (\xi) \to X^\mu (\xi) + \delta X^\mu (\xi) $.
Let us first decompose the deformation in its tangential and normal parts
with respect to the basis $\{ X{}' , n_i \}$,
\begin{equation}
\delta X  = \Phi X{}'  + \Phi^i n_i\,.
\end{equation}
Tangential deformations are reparametrizations
of the worldline.
All that we will need below is the tangential
variation of the infinitesimal proper time, 
\begin{equation}
\delta_\parallel d\tau = d\tau \Phi{}'\,.
\label{eq:pt}
\end{equation}
Now consider the normal part of the deformation. Its
effect on the geometry of a normal deformation
can be derived directly, or obtained by considering the
point-like limit of the expressions obtained in  \cite{Defo} for 
relativistic extended objects.
For the normal deformation of the 
infinitesimal proper time we have
\begin{equation}
\delta_\perp d\tau  =  d\tau \; k_i \; \Phi^i\,.
\label{eq:proper}
\end{equation}
For the normal deformation of the extrinsic curvature 
along the $i$-th normal we find
\[
\tilde\delta_\perp K^i = -  \tilde{\nabla}^2 \Phi^i 
+ k^i K_j \Phi^j \,.
\]
The symbol $\tilde\delta$ denotes
the normal rotation covariant deformation
operator defined in \cite{Defo}.
With these expressions we obtain that 
the normal variation of the mean curvature along the $i$-th normal is
\begin{equation}
\tilde\delta_\perp k^i = - 
\gamma^{-1} \tilde{\nabla}^2 \Phi^i 
- k^i k_j \Phi^j = \tilde{D}^2 \Phi^i -  k^i k_j \Phi^j\,,
\label{eq:mcurv}
\end{equation}
so that, using $\kappa_1 = k$, this gives
\begin{equation}
\delta_\perp \kappa_1 =    \hat{k}_i 
\left( \tilde{D}^2 \Phi^i 
- \kappa_1^2 \Phi^i \right)  \,.
\label{eq:mcurv1}
\end{equation}
In order to evaluate the variation of derivatives of the curvature we need also
the normal variation of the extrinsic twist,
\[
\tilde{\delta}_\perp \omega_{ij} =
 k_{j} \tilde{\nabla} \Phi_{i} -  k_{i} \tilde{\nabla} \Phi_{j} \,. 
\]
It follows that, for example,
\begin{equation}
\tilde\delta_\perp \left( \tilde\nabla k^i \right) =
- \gamma^{-1} \tilde\nabla^3 \Phi^i 
- k^2 \tilde\nabla \Phi^i - \Phi^j \tilde \nabla ( k^i k_j ) \,.
\label{eq:ddk}
\end{equation}

These expressions are sufficient to obtain the
variation of any geometrical invariant for the
worldline. However, as one considers higher FS
curvatures, the resulting expressions become 
increasingly unmanageable. For this reason it 
becomes desirable to develop an alternative approach where
the normal deformation is expanded with respect to
the FS basis as (we are suspending for the
remainder of this section the summation convention)
\begin{equation}
\delta_\perp X = \sum_i  \Psi_i \; \eta_i\,.
\end{equation}
Note that the normal components
of the deformation are related by a rotation, $\Phi^i = 
\sum R^i{}_j \Psi^j$.
Since $k_i = \kappa_1 
n_i \cdot \eta_1 $, we have that the normal variation of
the infinitesimal proper time takes the form
\begin{equation}
\delta_\perp d\tau = d\tau \; \kappa_1 \; \Psi_1 \,.
\end{equation}
It follows that the commutator between normal variation
and derivation by proper time is
\begin{equation}
\left[ \delta_\perp , {d \over d\tau}  \right] =  - \kappa_1 \;  \Psi_1 
 \,,
\end{equation}
so that in particular for the deformation of the tangent vector we have 
\[
\delta_\perp X' =
- \kappa_1 \Psi_1 X' + \left(\sum_i \Psi^i \eta_i \right)'
\,.
\]
Using the FS equations and rearranging the sum,
this expression reduces to 
 \begin{equation}
\delta_\perp X' = \sum_i \left[ \Psi^i{}' \eta_i 
+ \kappa_{i+1} \Psi^{i+1} \eta_{i} - \kappa_{i+1} \Psi^i \eta_{i+1} \right]\,.
\end{equation}
The projections of 
$\delta_\perp X' $ along the FS normals are  then
\begin{equation}
\eta^i \cdot \delta_\perp X' =  \Psi^i{}' 
+ \kappa_{i+1} \Psi^{i+1}  - \kappa_{i} \Psi^{i-1} \,,
\label{eq:xpp}
\end{equation}
where    $ \Psi^i=0$ if $i<1$ or $i>N$.
The corresponding projections of the variation of the
acceleration $X''$ is given by 

\begin{eqnarray}
\fl
\eta^i \cdot \delta_\perp X'' &=
\kappa_{i+2} \kappa_{i+1} \Psi^{i+2}
+ 2  \kappa_{i+1} \Psi^{i+1}{} ' + \kappa_{i+1}{}' \Psi^{i+1} 
+ \Psi^i {}'' - \left( \kappa_i^2 + \kappa_{i+1}^2 \right)
\Psi^i \nonumber \\
\fl
&-   2  \kappa_{i} \Psi^{i-1}{} ' - \kappa_{i}{}' \Psi^{i-1} 
+ \kappa_{i} \kappa_{i-1} \Psi^{i-2}\,,
\label{eq:xxpp}
\end{eqnarray}
where again $\Psi_i = 0 $ if $i<1$ or $i>N$.
Eq.(\ref{eq:xxpp}) follows from 
\[
\eta^i \cdot \delta_\perp X'' = (\eta^i \cdot \delta_\perp X' )'
- \eta^i{}' \cdot \delta_\perp X' -  \kappa_1 \Psi_1 (\eta^i\cdot X'')\,, 
\]
the FS equations and  (\ref{eq:xpp}).
Remarkably, this is all we need to evaluate the variations of the
FS curvatures. We will do this explicitly for 
$\kappa_1$, $\kappa_2$ and $\kappa_3$ (the latter only in a four-dimensional
background). The generalization 
to higher curvatures can be worked out along the same lines.

Taking a variation of the first of   (\ref{eq:frenet}) we have 
\begin{equation}
 \delta_\perp X'' = (\delta_\perp \kappa_1 ) \eta_1 
+ \kappa_1 \delta_\perp \eta_1\,.
\label{eq:X2} 
\end{equation}
Using the orthogonality
of $\eta_1$ and $\delta_\perp \eta_1 $ we obtain
$
\delta_\perp \kappa_1 = \eta_1 \cdot \delta_\perp X''
$,
and specializing (\ref{eq:xxpp}) to $i=1$, we find
\begin{equation}
\delta_\perp \kappa_1 = \kappa_3 \kappa_2 \Psi_3
+ 2 \kappa_2 \Psi_2{}' + \kappa_2' \Psi_2   
+ \Psi_1{}'' - \left( \kappa_1^2 + \kappa_2{}^2 \right) \Psi_1
\,.
\label{eq:perpkappa}
\end{equation}
This expression should be compared with (\ref{eq:mcurv1}), using
$\Phi^i = \sum R^i{}_j \Psi^j$. Note that the
normal variation of $\kappa_1$ involves at most three normal directions.

To evaluate the variation of the second 
FS curvature, we take a variation of the
second of  (\ref{eq:frenet})
and dot with
$\eta_{2}$. 
We obtain 
\[
\delta_\perp \kappa_2 = 
\kappa_1 \eta_2 \cdot \delta_\perp X' - \eta_2 \cdot
\delta_\perp \eta_1{}'\,.
\]
We can rewrite the second term on the right hand side as
\begin{eqnarray}
\eta_2 \cdot
\delta_\perp \eta_1{}' &=& 
- \kappa_1 \Psi_1 ( \eta_2 \cdot \eta_1{}' ) 
+ \eta_2 \cdot
( \delta_\perp \eta_1 )' \nonumber \\ 
&=&  \kappa_1 \kappa_2 \Psi_1  + 
(\eta_2 \cdot
 \delta_\perp \eta_1 )' - \eta_2{}' \cdot
\delta_\perp \eta_1  
\nonumber \\
&=& \kappa_1 \kappa_2 \Psi_1 
+ \left[ {1 \over \kappa_1}  \eta_2 \cdot
 \delta_\perp X'' \right]'
- { \kappa_2 \over \kappa_1} \eta_1 \cdot
 \delta_\perp X''
+ { \kappa_3 \over \kappa_1} \eta_3 \cdot
 \delta_\perp X''\,,
\nonumber
\end{eqnarray}
where we have used the second FS equation in the second line
and the third in the last line. 
$\delta_\perp \kappa_2$ is now expressed in terms of known quantities.
A little algebra gives 
\begin{eqnarray}
\fl
\delta_\perp \kappa_2 &= 
\kappa_1(\Psi_2{}' - 2\kappa_2 \Psi_1) 
\nonumber\\
\fl
&-  
\left\{ {1\over \kappa_1} \Big[ \Psi_2{}''- 2 \kappa_2 \Psi_1{}' - \kappa_2{}' \Psi_1  
- (\kappa_2^2 + \kappa_3^2)\Psi_2  + 2 \kappa_3 \Psi_3{}' + \kappa_3{}' \Psi_3 
+ \kappa_3 \kappa_4  \Psi_4 \Big]\right\} ' \nonumber \\
\fl
&-\,\,{\kappa_3\over \kappa_1} \,\,
\Big[  \kappa_2 \kappa_3 \Psi_1 
 - 2\kappa_3 \Psi_2{}'
 - \kappa_3' \Psi_2 + \Psi_3{}'' 
- (\kappa_1^2 + \kappa_3^2 + \kappa_4^2)\Psi_3 
\nonumber \\
\fl
&
\quad\quad \quad\quad\quad +  2\kappa_4 \Psi_4{}' + \kappa_4' \Psi_4 + \kappa_4 \kappa_5 \Psi_5 \Big]\,.
\label{eq:varkappa2}
\end{eqnarray}
Note that, whereas $\delta_\perp \kappa_1$ involves 
three normal directions, $\delta_\perp\kappa_2$ involves at most
five. Let us remark that an important simplification occurs 
in  this expression when we are in three 
dimensions and $\kappa_2$ 
is the top FS curvature ($\kappa_3 = \kappa_4 = \kappa_5
= 0$). The above expression reduces to
\begin{equation}
\fl
\delta_\perp \kappa_2 = 
\kappa_1(\Psi_2{}'  - 2\kappa_2 \Psi_1) -  
\left\{ {1\over \kappa_1} \Big[ 
\Psi_2{}''  - 2 \kappa_2 \Psi_1{}' - \kappa_2{}' \Psi_1
- \kappa_2^2 \Psi_2 \Big]\right\} '\,. 
\label{eq:varkappa2b}
\end{equation}

To complete the description of a particle moving in four dimensions one also requires 
an expression for the variation of the third FS curvature. Let $N=3$, so
that $\kappa_4 = \kappa_5
= 0$. We have 
\[
\delta_\perp \kappa_3 = 
{\kappa_2 \over \kappa_1}  \eta_3 \cdot \delta_\perp X'' - \eta_3 \cdot
\delta_\perp \eta_2{}'\,.
\]
Using the FS equations the second term
can be written as
\begin{equation}
\eta_3 \cdot
\delta_\perp \eta_2{}' 
= \kappa_1 \kappa_3 \Psi_1 
+ ( \eta_3 \cdot
\delta_\perp \eta_2{} )'\,.
\end{equation}
The second term in this expression can in turn 
be written as
\begin{eqnarray}
\eta_3 \cdot
\delta_\perp \eta_2 &=& {\kappa_1 \over \kappa_2}
\eta_3 \cdot \delta_\perp X' - {1 \over \kappa_2}
\eta_3 \cdot \delta_\perp \eta_1{}' \nonumber \\
&=& {\kappa_1 \over \kappa_2}
\eta_3 \cdot \delta_\perp X' - {1 \over \kappa_2}
\left( {1 \over \kappa_1} \eta_3 \cdot \delta_\perp X'' \right)'
+  {\kappa_3 \over \kappa_2 \kappa_1}
\eta_2 \cdot \delta_\perp X'' \,.
\nonumber
\end{eqnarray}
Using  (\ref{eq:xpp}) and (\ref{eq:xxpp}), 
we obtain
\begin{eqnarray}
\fl
\delta_\perp \kappa_3 &=&  - \kappa_1 \kappa_3 \Psi_1
+ {\kappa_2 \over \kappa_1} \Psi_3{}'' -  { \kappa_2 \kappa_3^2 \over 
\kappa_1}   \Psi_3  -
2 {\kappa_3 \kappa_2 \over \kappa_1}  \Psi_2{}' - {\kappa_2 \kappa_3{}'
\over \kappa_1}  \Psi_2 +  
 {\kappa_3  \over \kappa_1}  \kappa_2{}^2  \Psi_1 \nonumber \\
\fl
&-& \left\{ {\kappa_1 \over \kappa_2} ( \Psi_3{}' 
- \kappa_3 \Psi_2 ) \right\}'
+ \left\{  {1 \over \kappa_2}
\left[ {1 \over \kappa_1} \left( \Psi_3{}'' -  \kappa_3^2  \Psi_3  -
2 \kappa_3 \Psi_2{}' - \kappa_3{}' \Psi_2 
+ \kappa_3 \kappa_2 \Psi_1  \right)\right]' \right\}'
\nonumber \\ 
\fl
&-& \left\{ {\kappa_3 \over \kappa_2 \kappa_1}
\left[ \Psi_2{}'' - (\kappa_2^2 + \kappa_3^2 ) \Psi_2 + 2 \kappa_3 \Psi_3{}' + \kappa_3{}' \Psi_3 -
2 \kappa_2 \Psi_1{}' - \kappa_2{}' \Psi_1 \right] \right\}'\,.
\label{eq:varkappa3}
\end{eqnarray}
 
Let us remark that in the case of a curved background, 
 (\ref{eq:xxpp}) 
is modified according to
\begin{equation}
\eta_i \cdot \delta_\perp X'' \to \eta_i \cdot \delta_\perp X'' |_{flat} +
R_{\mu \alpha \nu \beta} X^{\mu}{}' X^\nu{}' \eta^\alpha{}_j \eta^\beta{}_i \Psi^j\,,  
\end{equation}
where the first term is given by  (\ref{eq:xxpp}) and 
$R_{\mu\alpha\nu\beta}$ denotes the
background Riemann tensor.
In order to obtain the normal variation of the FS curvatures 
in an arbitrary background one would have to make the appropriate
changes in the expressions we have derived. 

\section{Worldline invariants}

The action functionals for relativistic particles
we will consider satisfy various symmetry requirements.
First of all, we consider only actions which are invariant
under reparametrizations of the worldline of the
form $S = \int d\tau L $, with $L$ a scalar under reparametrizations, built out
of the geometrical quantities that characterize the
worldline, {\it i.e.} its curvatures.
This assumption implies that the first variation of the action can always be written as
\begin{equation}
\delta S  = \int d\tau  E_i \Phi^i +
 \int d\tau Q ' \,,
\label{eq:var}
\end{equation}
where $E_i$ denotes the Euler -Lagrange derivative,
and $Q$ is the Noether charge.
This specific form follows from the fact that
the tangential variation contributes only to the 
Noether charge. It is a simple consequence 
of the transformation properties of the action under reparametrizations.
Indeed using (\ref{eq:pt}) and the
fact that $\delta_\parallel L = \Phi L'$, 
the tangential part of the variation of the action is
a total derivative
\begin{equation}
\delta_\parallel S
= \int d\tau  (  L \Phi )' \,,
\label{eq:varpa}
\end{equation}
and does not contribute to the Euler-Lagrange part of the variation of the action.

The lowest order possibility is the relativistic massive particle, with a
constant Lagrangian 
$L_0 = - m$, and an action proportional to the proper time elapsed along the trajectory.
At higher order the Lagrangian will be 
a scalar constructed out of  higher derivatives of the embedding functions 
$ L = L ( X'' , X''', \cdots ) $.
However writing higher order Lagrangians this way 
makes it increasingly difficult 
to keep track of the possible scalars one can construct at each order and to establish 
which ones are independent. Our 
approach is to construct the Lagrangian out of
the $k^i$ and its (covariant) derivatives
$L = L ( k^i , \tilde{D} k^i , \tilde{D}^2 k^i , \cdots ) $ or,
equivalently, as we have established, as a function of the FS curvatures and their
derivatives.

The lowest non-trivial geometrical models depend only on $k^i$. Invariance under rotations of the 
normals implies that the Lagrangian must depend
on $k^i$ only in the combination $k$ or equivalently the first FS 
curvature,  $L  =  L (\kappa_1 ) $.
The next order allows a dependence on first derivatives
of the $k^i$, so that $L = L (k^i, \tilde{D} k^i )$.
Now what are the scalars we can construct at this
order? A moment of thought reveals that the building
blocks  must 
be $k , k' , \tilde{D} k^i \tilde{D} k_i  $.
There is one caveat: in the case of a 
three-dimensional ambient space, we have also the
independent combination given by the pseudo-scalar $\epsilon_{ij} k^i \tilde{D} k^j $.
In terms of the FS curvatures, at this order, we
have $ L = L (\kappa_1 , \kappa_1{}' , \kappa_2 )$,
and in the special case of a 
three-dimensional ambient space we have also the
possibility $ L = L (\overline{\kappa}_2 )$. At the next
 order, we will have  $ L = L (\kappa_1 , \kappa_1{}' , 
\kappa_1{}'', \kappa_2 , \kappa_2{}' , \kappa_3 )$.

\section{First-curvature models}

In this section we consider a relativistic particle
whose dynamics is determined by an action that
depends at most on the first FS curvature 
of its worldline, $L=  L ( \kappa_1 )$,
where $L$ is an arbitrary local function of its argument. 
We first derive the 
equations of motion and the Noether charges
associated with the underlying Poincar\'e invariance 
for this class of models. 
The normal
variation of the action is
\[
\delta_\perp S = \int d\tau \left[ k_i  
L \Phi^i + L_i \tilde\delta_\perp k^i
\right]\,,
\]
where we have used (\ref{eq:proper}), the Leibniz rule, and we have
introduced the quantity $
L_i =  d L / d k^i = L^* \hat{k}_i $.
An asterisk denotes a derivative of $L$ with respect to its argument, 
$\kappa_1$, and $\hat{k}^i = k^i / k$. We can now use (\ref{eq:mcurv}) to obtain 
\[
\delta_\perp S = \int d\tau \left[ k_i  
L \Phi^i
+ L_i  \tilde{D}^2 \Phi^i 
- L_i  k^i k_j \Phi^j \right]\,.
\]
Integrating by parts, we reduce this expression to 
\begin{equation}
\fl
\delta_\perp S = \int d\tau \left[  
\tilde{D}^2 L_i  + ( L     
-  L_j k^j ) k_i 
\right] \Phi^i 
+
\int d\tau \; 
\left[  L_i \tilde{D} 
\Phi^i  - \Phi^i \tilde{D} L_i  \right]{}'\,.
\label{eq:varpe}
\end{equation}

\subsection{Equations of motion}

By comparison of  (\ref{eq:varpe}) with  (\ref{eq:var}), we recognize by inspection the 
Euler-Lagrange derivative, $E^{(1)}{}_i$ (the superscript refers to the first-curvature
dependance), so that the equations of motion are 
\begin{equation}
E^{(1)}{}_i = \tilde{D}^2 L_i   + 
\left( L - L_j k^j \right) k_i = 0 \,.
\label{eq:eom12}
\end{equation}
These are non-linear ordinary differential equations of fourth
order in the embedding functions, or of second order
in the curvatures. It is instructive to 
compare this expression with the considerably more complicated 
corresponding equations (at least prior to gauge fixing) 
written directly in terms of the embedding functions. 

It is clear from the equations of motion that the
normal forms introduced in (\ref{eq:tis})
$T^{(N)}_{i_1\cdots i_{N-r}} = 0$ on shell for all 
$r>2$. 
 This is because the equations of
motion
express 
$\tilde{D}^2 L_i$ (and therefore
$\tilde{D}^2 k_i$) in terms of $k_i$. 
This implies that the third
FS curvature $\kappa_{3}$ will vanish.
The classical 
dynamics is confined to  at most a three dimensional Minkowski subspace.
Note that if Minkowski space is replaced by an arbitrary background, 
a non-diagonal term proportional to the 
Riemann curvature of the background gets introduced into (\ref{eq:eom12})
which spoils this confinement. 

An alternative way to see that $\kappa_3$ vanishes
is to use $L_i = L^* \hat{k}_i$ in the equations of motion. We have  
$L_i k^i = L^* \kappa_1$, and
the equations of motion assume the form
\[
L^* \; \tilde{D}^2 \hat{k}^i +
2 L^*{}' \; \tilde{D} \hat{k}^i
+ \left( L^*{}'' + L - L^* \kappa_1 \right) \hat{k}_i
= 0\,.
\]
We now project this equation into the three independent 
directions in the normal plane $\eta_1, \eta_2 , \eta_3$
and  then we find that they imply, along $\eta_3$,
\begin{equation}
E^{(1)}{}_3 = L^* \; \kappa_2 \; \kappa_3  = 0\,,
\label{eq:k3z}
\end{equation}
{\it i.e.} the vanishing of $\kappa_3$.
Moreover, we obtain along $\eta_2$, 
\begin{equation}
E^{(1)}{}_2 = - 2  L^*{}' \; \kappa_2  - L^* \; \kappa_2{}' = 0\,.
\label{eq:pl1k}
\end{equation}
This equation determines $\kappa_2$ in terms of $L^*$, or 
equivalently in terms of $\kappa_1$. It can be easily integrated to give
the relation
\begin{equation}
(L^* )^{2} \kappa_2 = \mbox{const.} 
\label{eq:l1k}
\end{equation}
As we will show below, this conservation law can 
be interpreted in terms of conservation of the spin.
The remaining projection of the equations of motion can 
be written as
\begin{equation}
E^{(1)}{}_1 = L^*{}'' + \left( L - L^* \kappa_1 \right) \kappa_1 - 
L^* \; \kappa_2^2  = 0 \,.
\label{eq:eomnew}
\end{equation}
We see that the second FS curvature contributes
to the ``driving force" in the equations of motion.
When $\kappa_2$ is expressed via   (\ref{eq:l1k}) in terms
of $\kappa_1$, the dynamical problem is reduced to
the motion of a one-dimensional fictitious particle with 
$\kappa_1$ as position variable.

A more systematic way to arrive at  (\ref{eq:k3z}), (\ref{eq:l1k}) , (\ref{eq:eomnew})
is desirable. This is possible by choosing the independent normal variations to be those 
projected along
the FS basis. Specifically, we have
\begin{eqnarray}
\fl
\delta S &=& \int d\tau\, \left\{L^* \delta_\perp \kappa_1 + L \kappa_1 \Psi^1 \right\}
+ \int d\tau\, (L \Phi)' \nonumber\\
\fl
&=& \int d\tau \, \Big [ L^*{}'' 
- (\kappa_1^2 + \kappa_2^2) L^* + L \kappa_1 \Big]\Psi_1 
 + \int d\tau\, \Big[ L^* \kappa_2{}' - (2 \kappa_2 L^*)'\Big] \Psi_2
\nonumber \\
\fl
&& + \int d\tau\, L^* \kappa_2\kappa_3\Psi_3
+ \int d\tau\, \left[ L\Phi + 2 \kappa_2 L^* \Psi_2\right]'\,,
\label{eq:delFS}
\end{eqnarray}
using the expression (\ref{eq:perpkappa}) for 
$\delta_\perp \kappa_1 $ in the normal variation of the action, and (\ref{eq:varpa}) for the tangential variation of the
action.
We immediately read off the equations of motion in the form
(\ref{eq:k3z}), (\ref{eq:l1k}) , (\ref{eq:eomnew}).

Let us consider some specific examples beginning with the degenerate ones. 
For the familiar
massive relativistic particle, $L = - m$, so that $L_i = 0$,
and the equations of motion reduce to $m \kappa_1 = 0$, {\it i.e.}
the vanishing of the particle acceleration. The linear correction to the relativistic particle,
$L = -m + \alpha \kappa_1$, where $\alpha$ is constant is also special. 
Now  (\ref{eq:l1k}) gives $\kappa_2$ is a constant and   (\ref{eq:eomnew}) reads
$m\kappa_1 + \alpha \kappa_2^2 =0$ so that $\kappa_1$ is also.
If $m=0$, then $\kappa_2=0$ and the motion is confined to a plane where 
the linear action is topological (with a vanishing 
Euler-Lagrange derivative almost everywhere) \cite{Ply}.
We will have more to say about this case later.
In all other cases, $L^*$ is not constant, and (\ref{eq:eomnew}) is of higher 
order. An interesting
non-polynomial model  
defined by $L = - \sqrt{k_0^2 - \kappa_1{}^2}$, with 
$k_0 =$ constant, describes a trajectory with a bounded acceleration $k_0$ 
\cite{Caiaetal} (see also \cite{Beem} for a discussion of bounded 
acceleration).
It has the  remarkable feature of having equations of motion linear
in the variable $L_i$. Indeed, since
$L_i =   k_i / L $, they take 
the form, $( \tilde{D}^2 - k_0^2) L^i  = 0$, 
which describes an upside down harmonic oscillator. 
Note that as $k\to k_0$ we have that $L^i \to \infty$. It is 
interesting to observe that in
 the corresponding theory described perturbatively 
by $ L =  - k_0 +  (1 / 2k_0) \kappa_1{}^2 \, + \cdots $,
the equations of motion are not harmonic at any finite order in the
expansion. 

It is well known that the equations of motion 
(\ref{eq:eom12}) are integrable; $\kappa_1$ is determined as a 
quadrature. For a relativistic particle, this was 
shown by  Nesterenko {\it et al.}   using a FS basis
\cite{Nesto}. 
In the next section we will offer an alternative proof which exploits
the conserved charges associated with the Poincar\'e symmetry of these models.
The advantage of this technique is that it can be 
exploited at higher orders to identify non-trivial first integrals of the 
equations of motion, a task which is not at all obvious generalizing the methods used
 after (\ref{eq:k3z}).

\subsection{Invariants}

The Noether charge $Q$ for $K(\kappa_1)$ models
is given by

\begin{equation}
\fl
 Q =   L \Phi + L_i \tilde{D} 
\Phi^i -
 \Phi^i \tilde{D} L_i  =  L\Phi 
+ L^* \Psi_1{}'- L^*{}'\Psi_1 + 2 L^* \kappa_2 \Psi_2\,.
\label{eq:qrig}
\end{equation}
The first expression is obtained by exploiting Eq.(\ref{eq:varpa}) and (\ref{eq:varpe}) 
and comparing with 
(\ref{eq:var}); for the second we use (\ref{eq:delFS}) instead of (\ref{eq:varpe})
($L^i = L^* \hat{k}^i$). 
From the Noether charge we obtain the 
conserved linear momentum by specializing the 
arbitrary deformation to a  constant infinitesimal translation,
$\delta X^\mu = \epsilon^\mu$, and setting
$ Q  = \epsilon_\mu P^\mu $.
With the help of  (\ref{eq:gwp}) or (\ref{eq:frenet})
this gives for the conserved linear momentum,
\begin{equation}
\fl
P^\mu = 
(L_i k^i - L ) X^\mu{}' - \left( \tilde{D} L^i 
\right) n^\mu{}_i =
(L^* \kappa_1  - L ) X^\mu{}' -  L^*{}' \eta^\mu_1
+ \kappa_2 L^* \eta^\mu{}_2 \,.
\label{eq:mom1b}
\end{equation}
Note that $P^\mu$ in general possesses non-vanishing 
projections along the first two FS normal directions.
For the relativistic free particle, $L = - m$,  we recover the 
familiar expression $P^\mu = m X^\mu{}'$. 
The momentum is
purely tangential to the worldline. As long as the
wordline remains timelike the momentum will be also. In general, however, 
when $L_i \neq 0$, there will be a non-trivial 
normal component. It follows that even if the trajectory is timelike, the momentum $P^\mu$ need not be. 
The invariant mass is
\begin{equation}
\fl
M^2  = - P^\mu P_\mu  = 
(L_i k^i - L )^2   - \tilde{D} L_i   \tilde{D} L^i 
= (L^* \kappa_1 - L )^2   - (L^*{}')^2 -  
(L^*)^2 \kappa_2^2  \,.
\label{eq:shell}
\end{equation}
The subtracted tachyonic terms  lower the mass. They vanish if and only if
$ \tilde{D} L^i = 0$. These are the {\it static}
 solutions
for this system. If $ (L - L_i k^i )^2   < 
\tilde{D} L_i   \tilde{D} L^i $, $P^\mu$ is 
spacelike. The occurence of tachyonic solutions is a well-known 
feature of higher derivative relativistic theories.
This expression
makes explicit that the mass is lowered both by variations
in the acceleration and by excursions away from the plane
determined by its velocity and its acceleration.
In the following we will assume that $M^2 \geq 0$.

In a similar way from the Noether charge we obtain the 
conserved angular momentum by specializing the 
arbitrary deformation to a spacetime  infinitesimal 
Lorentz transformation, 
$\delta X^\mu = \omega^\mu{}_\nu X^\nu$ with the constants 
$\omega_{\mu\nu} = - \omega_{\nu\mu}$, and setting
$ Q = \omega_{\mu\nu} M^{\mu\nu} $.
Using the first two of the FS equations 
 (\ref{eq:frenet}) this gives the conserved
angular momentum
\begin{equation}
M^{\mu\nu} =   P^{[\mu} X^{\nu]} +  N^{\mu\nu}\,,
\end{equation}
where 
\begin{equation}
N^{\mu\nu} =    L^i n^{[\mu}{}_i X^{\nu]}{}' 
= L^* \eta^{[\mu}_1  X^{\nu]}{}'\,.  
\end{equation}
Note that $M^{\mu\nu}$ is not generally of the simple orbital form unless 
$L_i = 0$, {\it i.e.} for the special case
of the standard relativistic particle with $L = -m$.
The bivector $N^{\mu\nu}$ is the origin of the spin of curvature-dependent 
particle models. However, $N^{\mu\nu}$ is not of the most general form: 
at this order there is no purely normal term of the form $m^{ij} n^{[\mu}{}_i n^{\nu]}{}_j$.
In order to evaluate the spin of a curvature-dependent
particle we define the Pauli-Lubanski pseudo-tensor ($N \geq 2$,
$M^2 > 0$)
\begin{equation}
\label{pau}
S_{\alpha_1 \cdots \alpha_{N-2}} =
{ 1 \over \sqrt{(N-2)!}}{ 1 \over\sqrt{M^2}}
\varepsilon_{\alpha_1\cdots\alpha_{N-2} \mu\rho\sigma}
P^{\mu} M^{\rho \sigma} \,.
\end{equation}
The Pauli-Lubanski pseudo-tensor picks up the non-orbital
part of the angular momentum, and
it is identified as the spin.
Since 
$\kappa_3 $ vanishes as a consequence of  (\ref{eq:k3z}),
we can, without loss of generality, set $N=2$ 
so that

\begin{equation}
S = 
{ 1 \over\sqrt{M^2}} 
\varepsilon_{\mu\rho\sigma}
P^\mu N^{\rho\sigma} = { 1 \over\sqrt{M^2}} 
\varepsilon_{\mu\rho\sigma}
  n^\mu{}_i n^\rho{}_j  X^\sigma{}' \left( \tilde{D} L^i 
\right) L^j\,.
\end{equation}
Note that the tangential projection of the linear momentum
plays no role at this order, therefore the spin is purely normal.
 The normal Levi-Civita density can be exploited to rewrite the 
Pauli-Lubanski pseudo-tensor in a more geometrical form,
\begin{equation}
S = { 1 \over\sqrt{M^2}} 
\varepsilon_{ij}  L^i \tilde{D} L^j =
{L^*{}^2 \over\sqrt{M^2}} 
\varepsilon_{ij} \hat{k}^i \tilde{D} \hat{k}^j 
= {L^*{}^2 \over \sqrt{M^2}} \overline\kappa_2 \,.
\label{eq:spn2}
\end{equation}
A necessary and sufficient condition 
for the Lagrangian $L(\kappa_1)$ to  be associated
with a spinless particle is the vanishing
of the second FS curvature, $\kappa_2$.
If we consider $L^i$ as a position variable, the first expression in 
(\ref{eq:spn2})
for the Pauli-Lubanski pseudo-tensor makes explicit that
it can be seen as the angular momentum associated
with invariance under normal rotations.
The Poincar\'e Casimir $S^2$ is given by 
\begin{equation}
M^2 S^2 = 
\left( \tilde{D} L^i 
\right)\left( \tilde{D} L_i 
\right) L^j L_j -  ( L_i  \tilde{D} L^i )^2 
= (L^*)^4 \kappa_2^2\,,
\label{eq:s3}
\end{equation}
which should be compared with  the equation of motion in the direction $\eta_2$, as given by
(\ref{eq:l1k}).

\subsection{Integrability}

In this section, we exploit the conserved quantities
$M^2$ and $S^2$, derived in the previous section, to provide
an elementary proof of the integrability of the 
model described by a Lagrangian of the form $L(\kappa_1 )$.
We have from the FS forms of (\ref{eq:shell}) and (\ref{eq:s3})
\begin{eqnarray}
M^2 &=&(L^* \kappa_1 - L )^2   - (L^*{}')^2 -  
(L^*)^2 \kappa_2^2  \nonumber\\
M^2 S^2 &=&   (L^*)^4 \kappa_2^2\,.
\end{eqnarray}
The later determines $\kappa_2$ as a function of $\kappa_1$ with
constant of proportionality given by the two Casimirs, $M^2$ and $S^2$.
Substituting into the former determines $\kappa_1$  as a quadrature,
\begin{equation}
M^2 =(L^* \kappa_1 - L )^2   - (L^*{}')^2 -  
{ M^2 S^2\over (L^*)^2}\,.
\label{eq:ene1}
\end{equation}
The analogue with the motion of a fictitious particle located at $L_i$
moving in a central potential should now be obvious.
To be explicit, note that its velocity can be decomposed into its 
projections along $L_i$ (and so $k_i$) and orthogonal to it ($(L^*)^2 = L_i L^i)$:
\begin{equation}  
\tilde D L_i  = D L^* \,\hat L_i + (\tilde D L_i)_\perp\,,
\end{equation}
where
$\tilde D L^j_\perp = (\delta^{jk} - \hat k^j \hat k^k)\tilde D L_k$
The right hand side of (\ref{eq:s3})
can alternatively be expressed in the form
$(L^*)^2 \tilde D L^j_\perp \tilde D L_j{}_\perp $.
We note that  the ``kinetic" term 
$
(\tilde D L^i)(\tilde D L_i)$ appearing in the mass
formula  (\ref{eq:shell}) can correspondingly be decomposed into its radial and 
tangential parts as 
\begin{equation}
(\tilde D L^i)(\tilde D L_i)
= |L^*|' {}^2 +  (\tilde D L^j_\perp)^2\,.
\end{equation}
We now exploit `angular momentum' conservation, 
 (\ref{eq:s3}) to eliminate  the tangential component in favor of a 
centrifugal potential. We obtain  (\ref{eq:ene1}).
If we identify $L^i$ with 
position in  $N$ dimensional Euclidean space, then 
 (\ref{eq:ene1}) describes the radial motion 
of a fictitious particle with unit mass moving 
in a central potential,
\begin{equation}
V(L^*)= - (L^i k_i - L)^2\,,
\end{equation}
where the right hand side should be seen as 
an implicit function of $\kappa_1$,  
with angular momentum $(N-1)S^2 M^2 $, 
and energy $-M^2$. 
We note that the potential
is negative everywhere.
In the degenerate case, $L= - m + \alpha \kappa_1$, 
from  (\ref{eq:ene1})
we obtain the mass-spin relation $M^2 = m^2 / (1 + S^2/\alpha^2)$. 
In all other cases 
the expression (\ref{eq:ene1}) can be integrated to give
\begin{equation}
\tau = \int { d L^* \over [
- V(L^*)
- M^2 +  M^2 S^2/ L^*{}^2 ]^{1/2} }\,.
\end{equation}
The simplest  model is 
described by  $L  = - m + \alpha \kappa_1{}^2 $. The potential takes the form 
$ V = - \left( \alpha \kappa_1{}^2  + m \right)^2\,.$
For the bounded acceleration model, we find
$V = - k_0^2 ( L^*{}^2 + 1) $.
The potential typically is negative. For a given real mass
$L^*$ is bounded from below by the centrifugal barrier 
provided by the spin. A trajectory bouncing off this  
barrier will have $L^*$ increase monotonically to infinity.
While this feature does translate into a divergence of $\kappa_1$ 
for all $\kappa_1{}^n$ models, $n\ge 2$, it does not always imply the 
divergence of $\kappa_1$ as the bounded accelation model clearly indicates.

\subsection{Non-flat background}

It is also instructive to approach this problem from 
an alternative point of view 
which does not rely (explicitly at least) on the global 
Poincar\'e invariance of the theory. 
Let us consider the antisymmetric normal 
tensor  $j^{ij}$ defined by 
\begin{equation}
j^{ij} = L^i \tilde{D} L^j - L^j \tilde{D} L^i\,.
\end{equation}
This is the 
angular-momentum tensor associated with the
configuration variable $L^i$ on the normal plane.
Note, however,  that $j^{ij}$ is not 
the angular momentum associated with the langrangian,  
treated as a function of $L_i$.
Its derivative  gives
\begin{equation}
\tilde{D} j^{i j} = 
\left( L^j \tilde{D}^2 L^i - 
L^i \tilde{D}^2 L^j \right) 
=   \left( L^l k_l - L \right) \left(
L^j k^i - 
L^i k^j \right)\,,
\end{equation}
where we have exploited 
the Euler-Lagrange equations in the form (\ref{eq:eom12}) to obtain the second equality. 
Since  $L^i \propto k^i$, we obtain $\tilde{D} j^{ij}=0$. Each component 
of this ``angular momentum" is conserved. 

We note that on a general spacetime background 
the Euler-Lagrange equations for the particle are modified
by the addition of a curvature dependent term as
\begin{equation}
\tilde{D}^2 L^i - ( L - L^j k_j) k^i 
+ R_{\mu\nu\rho\sigma} X^\mu{}' n^{\nu\,i} X^\rho{}' n^\sigma{}_j L^j = 0\,,
\label{eq:crv}
\end{equation}
where $R_{\mu\nu\rho\sigma} $ is the background Riemann tensor.
In particular, if the spacetime geometry has constant curvature, 
$R_{\mu\nu\rho\sigma}= c (g_{\mu\rho} g_{\nu\sigma} -
 g_{\mu\sigma} 
g_{\nu\rho})
$,
where $c$ is a constant, the curvature correction appearing in 
 (\ref{eq:crv}) is proportional to 
$L^i$. The $j^{ij}$ continue to be conserved.
This observation explains the integrability
of the first curvature particle models in spaces
of constant curvature found in Ref. \cite{Nesteita2}.

\subsection{Scale invariance}

Let us suppose that the particle model 
defined by $L(\kappa_1)$ 
is also invariant 
under a scale transformation, 
$\delta X^{\mu}= \varepsilon X^{\mu}$,
with $\varepsilon$ constant. 
It is simple to show from the expression (\ref{eq:qrig})
 for the
Noether charge that
the corresponding constant of the motion is given by 
$W = P^{\mu} X_{\mu}$.
When the equations of motion are 
satisfied, the conservation of $W$, $W'=0$, implies
the vanishing of the tangential component of the momentum,
the one-dimensional analogue of the vanishing of the trace
of the stress-energy tensor,
$ L - L_i k^i = L - L^* \kappa_1 =0 $.
The unique solution, 
up to a multiplicative constant, is the degenerate $L = \kappa_1$. 
Note that in the special case $N=1$, for a plane curve, there also exists the scale
invariant action given by the signed curvature
$\overline{\kappa_1}$, which is also a 
topological invariant, the winding number.

For this model, the momentum is
purely normal,
$
P^\mu = - (\tilde{D} \hat{k}^i ) n^\mu{}_i =
\kappa_2 \eta^\mu{}_2 $.
The Euler-Lagrange equations assume the
simple form
$
\tilde{D}^2  \hat{k}_i =0 $,
which implies $\kappa_2 = 0$, so that
 the momentum vanishes. Moreover 
 the spin vanishes as well and
the classical motion lies on a plane. But the Lagrangian 
$L=\kappa_1$ on a 
plane is simply the winding number if $\kappa_1$ is positive everywhere.
In any case, the Euler-Lagrange equations are identically satisfied almost 
everywhere. 

\subsection{Adding a derivative}

At the next order in derivatives the action might depend on $\kappa_1'$ or $\kappa_2$.
While at first sight, it might appear that these models 
possess a comparable level of complexity, we will see 
that this is not the case. 
Let us consider briefly a model depending  on the
first derivative of the first curvature, $\kappa_1{}'$,
for simplicity, with $L= L(\kappa_1{}')$. 
Again, the only variations give non-vacuous 
Euler-Lagrange equations are those along $\Psi_1,\Psi_2,\Psi_3$.
($L^{**} = dL / d \kappa_1{}' $): 
\begin{eqnarray}
L^{**}{}' \kappa_2^2 \kappa_3^2 &=& 0\,,
\\
 (L^{**}{}' )^4 \kappa_2^2 &=& \mbox{const.}\,,
\\
L^{**}{}''' -  ( L - L^{**} \kappa_1{}' + L^{**}{}' \kappa_1 ) \kappa_1 - L^{**}{}' \kappa_2^2 &=& 0\,.
\end{eqnarray}
The structure of these equations is 
very similar to that obtained with $\kappa_1$. As before,
$\kappa_3=0$ so that the motion is confined 
to a three-dimensional subspace of Minkowski spacetime.
The second equation is a conservation law, which expresses $\kappa_2$ in terms of derivatives
of $\kappa_1$. The main difference is in the first
equation, which is now of fourth order in derivatives
of $\kappa_1$. The invariant mass is given by
\begin{equation}
M^2 = \left( L^{**} \kappa_1{}' - L^{**}{}' \kappa_1 - L
\right)^2 - ( L^{**}{}'' )^2 - \kappa_2^2 \; ( L^{**}{}' )^2\,;
\label{eq:m2d}\end{equation}
the second Casimir is given by
$ M^2 S^2 = ( L^{**}{}' )^4 \; \kappa_2{}^2$
which permits $\kappa_2$ to be eliminated from Eq.(\ref{eq:m2d}).
The resulting expression does not, however, ever provide a quadrature.
Much more interesting are models depending on $\kappa_2$ to which we will now turn.

\section{Second-curvature models}

We extend now our considerations to a relativistic particle
whose dynamics is determined by an action depending on $\kappa_2$,  $L= L (\kappa_2 )$.
The tangential variation of the action is given by
(\ref{eq:varpa}). For  its normal variation
of the action we use (\ref{eq:varkappa2}) and the
Leibniz rule. There are non-vanishing variations along five of the FS normal directions
($L_* = d L / d \kappa_2$) 
\begin{eqnarray}
\fl
\delta_\perp S |_1 &=&
\int d\tau 
\left[ 2 \kappa_2 \left( { L_*{}' \over \kappa_1 }\right)'
+  {\kappa_2{}' \over \kappa_1} L_*{}'
+ L \kappa_1 
- L_* {\kappa_2 \over \kappa_1} ( 2\kappa_1^2 
+ \kappa_3^2 ) \right] \Psi_1
\nonumber \\
\fl
&+& \int d\tau  \left[
2 \; {\kappa_2 \over \kappa_1} L_* \Psi_1{}' +
{\kappa_2{}' \over \kappa_1} L_* \Psi_1
- 2 \; {\kappa_2 \over \kappa_1} L_*{}' \Psi_1 \right]'
\\
\fl
\delta_\perp S |_2&=&
\int d\tau 
\left[ \left( { L_*{}' \over \kappa_1} \right)''
 -  { L_*{}' \over \kappa_1}  (\kappa_2^2
+ \kappa_3^2 ) 
- ( L_* \kappa_1 )' - 2 \left( { L_* \kappa_3^2 \over \kappa_1}  \right)'
+   { L_* \kappa_3 \over \kappa_1}  \kappa_3{}' \right] \Psi_2
\nonumber \\
\fl
&+& \int d\tau  \left[ 
 - { L_* \over \kappa_1} \Psi_2{}''
+ { L_*{}' \over \kappa_1}  \Psi_2{}'
+  { L_* \over \kappa_1} ( 3 \kappa_3^2
+ \kappa_1^2 
+ \kappa_2^2 ) \Psi_2 
- \left( { L_*{}' \over \kappa_1} \right)' \Psi_2
 \right]'
\\
\fl
\delta_\perp S |_3&=&
\int d\tau 
\left[  -  \left( {L_* \kappa_3 \over \kappa_1} \right)''
- 2 \left( {L_*{}' \kappa_3 \over \kappa_1}\right)' 
+ L_*{}' {\kappa_3{}' \over \kappa_1}
 + { L_* \kappa_3 \over \kappa_1} (\kappa_1^2 + \kappa_3^2 
+ \kappa_4^2 ) 
\right] \Psi_3
\nonumber \\
\fl
&+& \int d\tau \left[ 
- 3 { L_* \kappa_3 \over \kappa_1}  \Psi_3{}'
+ \left( { L_* \kappa_3 \over \kappa_1} \right)' \Psi_3
- { L_* \kappa_3{}' \over \kappa_1} \Psi_3
+ 2 { L_*{}' \kappa_3 \over \kappa_1} \Psi_3 \right]'
\\
\fl
\delta_\perp S |_4&=&
\int d\tau 
\left[  3  { L_*{}' \kappa_3 \kappa_4 \over \kappa_1} 
+ 2  L_* \left( {\kappa_3 \kappa_4 \over \kappa_1} \right)'
- { L_*\kappa_3 \kappa_4{}' \over \kappa_1}  \right] \Psi_4
- \int d\tau  \left[ 3 { L_* \kappa_3 \kappa_4 \over \kappa_1}   \Psi_4 \right]'
\\
\fl
\delta_\perp S |_5 &=& - \int d\tau {L_*  \kappa_3 \kappa_4 \kappa_5 \over
\kappa_1 }
\Psi_5
\end{eqnarray}
The corresponding Euler-Lagrange derivatives 
can then be read off by 
discarding total derivatives: For $\Psi_5$ we have 
\begin{equation}
 E^{(2)}{}_5 = - { L_*\kappa_3 \kappa_4 \kappa_5 \over  
\kappa_1 } = 0 \,.
\end{equation}
This equation of motion sets $\kappa_5 = 0$. Motion will be restricted to (at most) a
5-dimensional subspace. 
For $\Psi_4$, we have
\begin{equation}
  E^{(2)}{}_4 = 3 { L_*{}' \over  \kappa_1 } \kappa_3 \kappa_4
+ 2 L_* \kappa_4 ( {\kappa_3 \over \kappa_1 } )'
+ { L_*  \over \kappa_1 } \kappa_3 \kappa_4{}' = 0\,.
\end{equation}
This equation of motion  can be easily integrated to give 
$ L_*{}^3  \kappa_3^2 \kappa_4/ \kappa_1{}^2 
= \mbox{const.}$,
which determines $\kappa_4 $ in terms of the lower curvatures. 
Note that this is the analogue of 
(\ref{eq:l1k}) for first curvature models, although in this case
it does not appear to be derivable from a conserved charge.

The remaining three Euler-Lagrange derivatives are 
\begin{eqnarray}
\fl
E^{(2)}{}_3 &=&  -  3 L_*{}'' {\kappa_3 \over \kappa_1}
-4 L_*{}' \left({\kappa_3 \over \kappa_1}\right)' 
+ L_*{}' {\kappa_3{}' \over \kappa_1}
- L_* \left({\kappa_3 \over \kappa_1}\right)''  \nonumber \\
\fl
 &+& { L_* \kappa_3 \over \kappa_1} (\kappa_1^2 + \kappa_3^2 
+ \kappa_4^2 ) = 0\,, \\
\fl
E^{(2)}{}_2 &=& \left( { L_*{}' \over \kappa_1} \right)''
 - \left( { L_*{}' \over \kappa_1} \right) (\kappa_2^2
+ \kappa_3^2 ) 
- ( L_* \kappa_1 )' - 2 \left( { L_* \kappa_3^2 \over \kappa_1}  \right)'
+  \left( { L_* \kappa_3 \over \kappa_1} \right) \kappa_3{}' = 0\,, 
\label{eq:ek2a}\\
\fl
E^{(2)}{}_1 &=&  2 \kappa_2 \left( {L_*{}' \over \kappa_1} \right)'
+  {\kappa_2{}' \over \kappa_1} L_*{}'
+ L \kappa_1 - L_* {\kappa_2 \over \kappa_1} ( 2\kappa_1^2 
+ \kappa_3^2 ) = 0\,.
\label{eq:ek2b}
\end{eqnarray}
Let us remark that these three higher-order non-linear
coupled ODEs do not appear to be tractable in general.

The corresponding Noether charge
is given by
\begin{eqnarray}
\fl
Q &=&  L \Phi  + 2 \; {\kappa_2 \over \kappa_1} L_* \Psi_1{}' +
{\kappa_2{}' \over \kappa_1} L_* \Psi_1
- 2 \; {\kappa_2 \over \kappa_1} L_*{}' \Psi_1 
- \left( { L_* \over \kappa_1} \right) \Psi_2{}''
\nonumber \\
\fl
&& + { L_*{}' \over \kappa_1}  \Psi_2{}'
- \left( { L_*{}' \over \kappa_1} \right)' \Psi_2
+
\left( { L_* \over \kappa_1} \right) ( 3 \kappa_3^2
+ \kappa_1^2 
+ \kappa_2^2 ) \Psi_2 
\nonumber \\
\fl
&-&  3 { L_* \kappa_3 \over \kappa_1}  \Psi_3{}'
+ \left( { L_* \kappa_3 \over \kappa_1} \right)' \Psi_3
- { L_* \kappa_3{}' \over \kappa_1} \Psi_3
+ 2 { L_*{}' \kappa_3 \over \kappa_1} \Psi_3 -
 3 { L_* \kappa_3 \kappa_4 \over \kappa_1}   \Psi_4\,. 
\end{eqnarray}
This permits us to write down the conserved momentum as
\begin{eqnarray}
\fl
P^\mu &=& (\kappa_2 L_* - L ) X^\mu{}' 
-  {\kappa_2 \over \kappa_1} (L_* )' \eta^\mu_1 
+ \Bigg[  - \left( { L_*{}' \over \kappa_1} \right)'
+   { L_* \over \kappa_1} ( \kappa_1^2  + \kappa_3^2 )\Bigg] \eta^\mu{}_2
\nonumber \\
\fl
&+& \Bigg[ \left( { L_* \kappa_3 \over \kappa_1} \right)' 
+ { L_*{}' \kappa_3 \over \kappa_1} \Bigg]\eta^\mu{}_3
-  {L_* \kappa_3 \kappa_4 \over \kappa_1 } \,\eta^\mu{}_4\,, 
\end{eqnarray}
which now possesses components along the first four FS
normals. Let us emphasize the similarity of the first two
components, along $X'$ and $\eta_1$ with the conserved 
momentum for first curvature models, given by (\ref{eq:mom1b}). 
The non-trivial invariant $M^2 = - P^\mu P_\mu$ can be written down
immediately, but it is not particularly illuminating.
The angular momentum is given by 
$M^{\mu\nu} = P^{[\mu} X^{\nu ]} + N^{\mu\nu} $
where the non-orbital part is given by
\begin{equation}
N^{\mu\nu} = 
{L_*{}' \over \kappa_1 } 
\eta^{[\mu}{}_2 X^{\nu ]}{}' - 
{L_* \kappa_3 \over \kappa_1 } 
\eta^{[\mu}{}_3 X^{\nu ]}{}' +
L_* \eta^{[\mu}{}_1  \eta^{\nu ]}{}_2 \,,
\label{eq:nk2}
\end{equation} 
which allows one to obtain the spin pseudo-tensor for this
class of models.

It is straightforward  to combine these results
 with the one obtained in Sect. 5, to obtain equations of motion and
conserved quantities for models of the form $L = L (\kappa_1 , 
\kappa_2 )$. One has only to be careful not to double count 
the contribution from the normal variation of the infinitesimal 
proper time and from the parallel variation of the action.
The special case of models linear in the FS curvatures
will be considered elsewhere.

\subsection{Integrability in three dimensions}
\label{i3d}

In general, as we have remarked, the system of equations determining the motion appears 
to be intractable. Let us specialize to a three-dimensional ambient space,
so that $\kappa_3 = \kappa_4 = \kappa_5 = 0$.
Moreover, we specialize  to the case of an action quadratic
in the second curvature, $L = -m +  (1/2) \alpha\kappa_2{}^2$. The case of 
an arbitrary $L$ can be treated along the same lines. Only the
degenerate case linear in the second curvature needs special
treatment (see below).
Surprisingly, we find that this model is integrable. Let us
show how this comes about.
The two surviving Euler-Lagrange equations (\ref{eq:ek2a}), (\ref{eq:ek2b})
reduce to 
\begin{eqnarray}
{E^{(2)}{}_2\over \alpha} &=& \left( { \kappa_2{}' \over \kappa_1} \right)''
 - \left( { \kappa_2{}' \over \kappa_1} \right) \kappa_2^2 
- ( \kappa_2 \kappa_1 )' = 0\,, \\
{E^{(2)}{}_1 \over\alpha} &=&  2 \kappa_2 \left({\kappa_2{}' \over \kappa_1} \right)'
+  {(\kappa_2{}')^2 \over \kappa_1} 
- {3 \over 2} \kappa_2{}^2 \kappa_1 - {m\over \alpha} \kappa_1= 0 \,.
\label{eq:e3k2}
\end{eqnarray}
The momentum takes the form
\begin{equation}
{P^\mu\over \alpha} =\left({m\over \alpha} + {\kappa_2^2 \over 2}\right)X^\mu{}' 
-  {\kappa_2 \kappa_2{}' \over \kappa_1} \, \eta^\mu_1 
- \Big[   \left( { \kappa_2' \over \kappa_1} \right)'
-   \kappa_1\kappa_2 \Big] \eta^\mu{}_2\,,
\label{eq:mk23}
\end{equation}
so that the invariant mass is 
\begin{equation}
- {M^2 \over \alpha^2}  = 
 \Bigg[ \left( { \kappa_2' \over \kappa_1} \right)'
-  \kappa_1\kappa_2 \Bigg]^2 - \left( {m \over \alpha} + {\kappa_2^2 \over
2} \right)^2
+  {\kappa_2^2 \kappa_2{}'^2 \over \kappa_1^2} \,.
\label{eq:k2m}
\end{equation}
We assume that $M^2 > 0$.
Now, from (\ref{eq:nk2}), (\ref{eq:mk23}), one obtains that
the spin pseudo-scalar is
\begin{equation}
\sqrt{M^2}S = \alpha \kappa_2 \left[ 
m + {\alpha \over 2} \kappa_2^2  
- \alpha \left( { \kappa_2' \over \kappa_1} \right)^2 \right]\,.
\label{eq:k2s}
\end{equation}
This can be rewritten as
\begin{equation}
\alpha \kappa_2{}'{}^2 + \kappa_1{}^2 \left( {\sqrt{M^2} S \over
\alpha \kappa_2} -
{\alpha \over 2} \kappa_2{}^2 - m \right) = 0\,.
\label{eq:into}
\end{equation}
We can now exploit (\ref{eq:e3k2}), (\ref{eq:k2m}), to obtain
\begin{equation}
\kappa_1{}^2 = {\alpha^2 \kappa_2{}^4 ( 4 \kappa_2 \sqrt{M^2} S - 4 M^2
+ 4 m^2 - \alpha^2
\kappa_2{}^4 )
\over ( \sqrt{M^2} S - \alpha \kappa_2{}^3 )^2 } \,.
\label{eq:poly}
\end{equation}
This expression can now be substituted into  (\ref{eq:into}) to
provide a quadrature for $\kappa_2$.

Note the equation of motion along $\eta_1$ (\ref{eq:ek2b}) can be interpreted
in terms of conservation of the spin, in the sense that
$ \sqrt{M^2} S' = - ( \kappa_2{}' / \kappa_1 ) E^{(2)}{}_1 $.
Recall that for the first curvature models it was the equation
of motion along $\eta_2$ which was related to spin conservation.
It follows, unfortunately, that our result does not appear to
extend to the more general case of models that depend on both
curvatures, $L = L (\kappa_1 , \kappa_2 )$.
Two special cases of interest are, however, integrable: (i) 
$L(\kappa_1)$ with a linear term in $\kappa_2$ added; (ii) $L= L(\kappa_2)$
in three dimensions with a linear term in $\kappa_1$ added.

\subsection{Scale invariance}

Invariant 
under scale transformations, 
implies as before
the vanishing of the tangential component of the momentum,
$  L - L_* \kappa_2 =0 $. As we found earlier for first curvature
models,
the unique solution, up to a multiplicative constant, is the degenerate $L = \kappa_2$. 

The equation of motion along $\eta_1$, (\ref{eq:ek2b}) reduces to
$
\kappa_2  (\kappa_1{}^2 + \kappa_3{}^2 ) /\kappa_1= 0$,
which implies $\kappa_2 = 0$. Using this in (\ref{eq:ek2a})
gives $\kappa_1 = $ const. 
The momentum is
purely normal,
$P^\mu = \kappa_1 \eta^\mu{}_2$ and
$M^2 = - \kappa_1{}^2$.
The motion is tachionic. Moreover the spin vanishes.

\subsection{Kuznetsov-Plyushchay model}

Althought the generic case $L = L ( \kappa_1 ,
\kappa_2 )$ appears to be intractable, there is in addition to the 
two integrable cases mentioned at the end of (\ref{i3d}), 
a large class of models in three dimensions 
which is also. These are the models 
defined by 
\begin{equation}
L = - m - \kappa_1 f ( x )\,, \;\;\;\;\;\;\;\;
x = \kappa_1 /\kappa_2 \,,
\label{eq:mkp}
\end{equation}
where $f$ is an arbitrary function of its argument. These models were
introduced by Kuznetsov and Plyushchay in \cite{KP1} who 
exploited a Hamiltonian approach.

The interest of this class of models lies in the
fact that they have simple solutions and that it
is easy to isolate the non-tachyonic solutions.  
It follows from the form (\ref{eq:mkp}) that we have
$ L^* = - f - x L_* $.
It is a straightforward to show that 
the momentum takes the remarkably simple
form
$ P^\mu = m (X^\mu{}' - x \eta^\mu{}_2)$, so that, in agreement with \cite{KP1}, 
$M^2 = m^2 ( 1 - x^2 )$.
From this expression, one obtains non-tachyonic
solutions by requiring $ x^2 \leq 1$. We emphasize
that this is independent of the function $f$. 
The spin also takes the  simple form
$ S = - m f (x) / \sqrt{M^2}$,
from which it follows that
$ f^2 (x) = S^2 ( 1 - x^2 ) $.
The solutions of this model
are given by constant $x$, as one can also
verify directly from the equations of motion.

\section{Conclusions}

We have examined the simplest relativistic object, a point particle,
described by a local geometrical action by 
developing a theory of deformations tailored to the FS basis
adapted to the particle worldline, and exploiting the 
Noether charges associated with Poincar\'e invariance.
In particular, we demonstrated explicitly how these techniques 
may be applied not only to the well known class of models
depending  on the first FS curvature of the worldline,
but also we break new ground using them to reveal remarkable 
properties of models depending on the second FS
curvature. We are able to show that the
latter models are also integrable in three dimensions. This is particularly
surprising when we consider that the Euler-Lagrange equations 
describing the dynamics are sixth order non-linear coupled ODEs.
Our work suggests a novel approach to the Hamiltonian
analysis for these models, as well as a point of departure for 
examining higher order relativistic extended objects.

\ack

RC is partially supported by CONACyT under grant 32187-E.
JG is partially supported by CONACyT 
under grant 32307-E and DGAPA at UNAM under grant IN119799.

\end{document}